\documentclass[%
 reprint,
superscriptaddress,
showpacs,preprintnumbers,
 amsmath,amssymb,
 aps,
showkeys,
floatfix,
]{revtex4-1}

\usepackage{graphicx}% Include figure files
\usepackage{dcolumn}% Align table columns on decimal point
\usepackage{bm}% bold math
\usepackage{ bbold }

\usepackage{color}
\usepackage{comment}
\usepackage{natbib}

\begin{document}

\preprint{Preprint}

\title{The Ba$_{0.6}$K$_{0.4}$Fe$_2$As$_2$ superconducting four-gap temperature evolution: 
a multi-band Chebyshev-BdG approach}% Force line breaks with 
%\thanks{Alternative title: High coupling ratios $2\Delta/k_B T$ in weak coupling BdG theory: application to iron-based superconductors.}%

\author{David M\"{o}ckli
}%
\affiliation{%
Theoretische Physik, ETH-Z\"{u}rich, 8093 Z\"{u}rich, Switzerland.\\
}
\affiliation{%
Instituto de F\'{\i}sica, Universidade Federal Fluminense, Niter\'oi, RJ, 24210-340, Brazil.\\
}

\email{david.moeckli@itp.phys.ethz.ch}

\author{E. V. L. de Mello}
 %\altaffiliation{Instituto de F\'{\i}sica, U, Niteroi, RJ, 24210-3niversidade Federal Fluminense40, Brazil.}%Lines break automatically or can be forced with \\
\affiliation{%
Instituto de F\'{\i}sica, Universidade Federal Fluminense, Niter\'oi, RJ, 24210-340, Brazil.\\
}

\date{\today}% It is always \today, today,

\begin{abstract}
We generalize the Chebyshev-Bogoliubov-deGennes method to treat multi-band systems
to address the temperature dependence of the superconducting (SC) gaps of iron based superconductors. 
Four  SC gaps associated with different electron and hole pockets of optimally doped
Ba$_{0.6}$K$_{0.4}$Fe$_2$As$_2$ were clearly identified by angle resolved photo-emission spectroscopy. The few approaches that reproduces with success this gap structure are 
based on strong-coupling theories and required many adjustable parameters.
We show that an approach with a redistribution of electron population between the hole and electron pockets $\nu$ with evolving temperature reproduces the different coupling ratios $2\Delta^{\nu}(0)/k_{\rm B} T_c$ in these materials. We define the values that fit the four zero temperature gaps $\Delta^{\nu}(0)$
and after that all $\Delta^{\nu}(T)$ is obtained without any additional parameter.
%\begin{description}
%\pacs{74.70.Xa}
%\end{description}
\end{abstract}

%Use showkeys class option if keyword
\maketitle

%\tableofcontents

\section{Introduction}

Iron based high-$T_c$ superconductors (FeSCs) have been intensely studied, but there are still many 
fundamental open question concerning their mechanism of pairing. The 
main difficulty stems from their multi-band structure with indications
of hybridisation among them. In this complex context,
the strength of the electron-electron correlation is  an issue of debate and  this is one of the main points addressed here. 
The temperature dependence of the SC gap often indicates the coupling regime of Cooper pairs. 
Conventional BCS superconductors are characterized by a weak-coupling strength ratio of $2\Delta(0)/k_{\rm B} T_c\approx 3.52$, whereas strongly correlated superconductors display much higher values.
%In cuprate superconductors $2\Delta(0)/k_{\rm B} T_c\approx 8$ were reported, which is interpreted as a signature of the strong correlations present in these materials \cite{Gomes2007}. 
In the
case of Ba$_{0.6}$K$_{0.4}$Fe$_2$As$_2$
a ratio of $\approx7.5$ is observed in the $\alpha$, $\gamma$ and $\delta$ pockets, while a ratio of $\approx 3.7$ is seen in the $\beta$ pocket at the Fermi surface  \cite{Ding2008,Terashima2009,Nakayama2009}. However, it is still unclear whether the strenght of the electron-electron interaction in the FeSCs is the cause of such high coupling ratios. 

Despite the clear distinction of four different SC band-gaps \cite{Nakayama2009}, optimally doped Ba$_{0.6}$K$_{0.4}$Fe$_2$As$_2$ displays a single critical temperature $T_c=37$ K. According to a well known result, this is a signal that the bands have an inter-dependent dynamics \cite{Suhl1959}. The different coupling strength ratios $2\Delta^{\nu}(0)/k_{\rm B} T_c$  in the FeSCs are frequently interpreted as a coexistence of different coupling regimes \cite{Richard2011,Richard2015}. It is not
intuitive that bands originated from different iron $d$-orbitals would possess distinct regimes.

To deal with this problem we develop a generalization of 
the Chebyshev-Bogoliubov-deGennes (CBdG) method \cite{Covaci2010} to treat multi-band systems. This
 weak-coupling mean-field approach reproduces the four-gap structure in Ba$_{0.6}$K$_{0.4}$Fe$_2$As$_2$, by allowing the electron population to redistribute among the bands with evolving temperature. We show that bands with monotonically varying electron population with temperature can generate high coupling strength ratios $2\Delta^{\nu}(0)/k_{\rm B} T_c$ in multi-band systems, as would be expected in strong-coupling systems. After fitting the values of $\Delta^{\nu}(0)$, our theory reproduces the temperature evolution of the four $\Delta^{\nu}(T)$ exactly as in BCS, without introducing any new parameters.  Then we show that a typical gap dependence $\Delta(T)$ is obtained by following a geodesic of constant chemical potential $\mu$ on the surface of $\Delta(\mu,T)$. The main point tackled in this paper shows that the four-gap structure of Ba$_{0.6}$K$_{0.4}$Fe$_2$As$_2$ is reproduced by geodesics on the surface of $\Delta(\mu,T)$, where $\mu = \mu(T)$. Whereas $\mu(T)$ varies monotonically for each band (and hence also the band  population) the total density of the system is constant.

\section{The model}

While in the cuprates the low-energy physical properties are captured by a single band, it is generally believed that
a minimal model for the FeSCs must include all five $3d$ orbitals of iron \cite{Hirschfeld2011}.
It has been shown that the charge excitations in different orbitals can be decoupled, so that it can be effectively described by a collection of doped Hubbard-like Hamiltonians, each with a different electron population \cite{DeMedici2014}.
As mentioned,
angle resolved photo-emission spectroscopy (ARPES)
on Ba$_{0.6}$K$_{0.4}$Fe$_2$As$_2$ measures four SC bands labeled $\alpha$, $\beta$, $\gamma$ and $\delta$ at the Fermi surface \cite{Nakayama2009}.

Here we address only these four bands because
our main goal is to reproduce their SC gap temperature dependence and, in doing so,
obtain some insights on the paring mechanism.
To set the CBdG method for the Ba$_{0.6}$K$_{0.4}$Fe$_2$As$_2$, we model each SC band by a square lattice, where each site represents a single decoupled orbital of Fe. Therefore, we model the Fe square lattice by four square lattices (sheets), each corresponding to a single band composed of decoupled $3d$ orbitals. This brings about a scenario where different pairing potentials may coexist. Inter-bands scattering is included as a single-particle scattering among these sheets. 

\begin{table}[]
\caption{\label{tab:parameters}Effective hopping parameters $t_1$ (nearest neighbor) and $t_2$ (next-nearest neighbor) consistent with APRES band dispersion fit \cite{Richard2011}. $U_0$ is the on-site $s$-wave pairing potential used  in the
BdG calculations of $\Delta^\nu_{\mathbf{i}}$.
%, $U_1$ is the nearest neighbor pairing potential, and $U_2$ is the next-nearest pairing potential.
}

\begin{ruledtabular}
\begin{tabular}{lllll}
{\bf (meV)} & { $\boldsymbol{\alpha}$} & { \boldsymbol{$\beta$}} & {\boldsymbol{$\gamma$}} & {\boldsymbol{$\delta$}} \\
$t_1$ & 160 & 13 & 380 & 380 \\
$t_2$ & -52 & 42 & 800 & 800 \\
$U_0$ & 227 & 52 & 1013 & 982 \\
% $U_1$ &  &  &  &  \\
% $U_2$ &  &  &  & 
\end{tabular}
\end{ruledtabular}
\end{table}

Our multi-band Bogoliubov-deGennes (BdG) Hamiltonian \cite{Mockli2015} is 
then composed by two parts, an intra-band and an inter-band component: $\mathcal{H}=\mathcal{H}_{\mathrm{intra}}+\mathcal{H}_{\mathrm{inter}}$, where
\begin{equation}
\begin{split}
\mathcal{H}_{\mathrm{intra}}&=\sum_{\langle\mathbf{ij}\rangle,\nu,\sigma}t_{\mathbf{ij}}^{\nu}\,c^\dag_{\mathbf{i}\nu\sigma}c_{\mathbf{i}\nu\sigma}-\sum_{\mathbf{i,\nu,\sigma}}\mu_\nu(T)\, c^\dag_{\mathbf{i}\nu\sigma}c_{\mathbf{i}\nu\sigma} \\
&+\sum_{\mathbf{i},\nu}\left(\Delta_{\mathbf{i}}^\nu \,c^\dag_{\mathbf{i}\nu\uparrow}c^\dag_{\mathbf{i}\nu\downarrow}+\mathrm{h.c.} \right )\\
&+\sum_{\langle\mathbf{ij}\rangle,\nu}\left(\Delta_{\mathbf{ij}}^\nu \,c^\dag_{\mathbf{i}\nu\uparrow}c^\dag_{\mathbf{j}\nu\downarrow}+\mathrm{h.c.} \right ),
\end{split}
\end{equation}
with
\begin{equation}
\Delta_\mathbf{i}^\nu= U_\mathbf{i}^\nu\langle c_{\mathbf{i}\nu\downarrow}c_{\mathbf{i}\nu\uparrow}\rangle,\quad\mbox{and}\quad
\Delta_\mathbf{ij}^\nu= U_\mathbf{ij}^\nu\langle c_{\mathbf{j}\nu\downarrow}c_{\mathbf{i}\nu\uparrow}\rangle,
\end{equation}
describes the intra-band dynamics, where the band index $\nu$ runs over the $\alpha$, $\beta$, $\gamma$ and $\delta$ band. The $t_{\mathbf{ij}}^{\nu}$ are intra-band hoppings between lattice sites $\mathbf{i}$ and $\mathbf{j}$ up to second nearest neighbors as derived by ARPES band dispersion tight-binding fit \cite{Richard2011}; see table \ref{tab:parameters}. 
The temperature dependent band chemical potential $\mu_\nu(T)$ allows for self-consistent regulation of the band fillings with temperature evolution.
The local gap amplitude $\Delta_\mathbf{i}^\nu$ realizes a constant $s$-wave gap. The non-local gap amplitudes $\Delta_\mathbf{ij}^\nu$, which can be nearest or next-nearest neighbors, realize unconventional gap symmetries.

Single particle inter-sheet scattering is usually described by \cite{Zhang2014,Balatsky2006}
\begin{equation}
\mathcal{H}_\mathrm{inter}=\sum_{\langle\mathbf{ij}\rangle,\mu\neq\nu,\sigma} \left(V_{\mathbf{ij}}^{\mu\nu}c^\dag_{\mathbf{i}\mu\sigma}c_{\mathbf{j}\nu\sigma}+\mathrm{h.c.} \right ), 
\end{equation}
where $V_{\mathbf{ij}}^{\mu\nu}$ is a non-local (nearest-neighbor) scattering potential among the bands and causes the multi-gaps to vanish at a common critical temperature $T_c=37$ K \cite{Mockli2015,Suhl1959}. The scattering strength is strong between bands close in momentum space (low momentum transfer), and weak for distant bands (high momentum transfer) \cite{Zhang2014}. Taking this into account,
we use $V^{\alpha\beta}=V^{\gamma\delta}=100$ meV and neglect all other distant bands scatterings.

\section{The multi-band CB$\rm d$G method}

As it is common to BdG method, we determine self consistently
the gap-functions $\Delta_\mathbf{i}^\nu$ and $\Delta_\mathbf{ij}^\nu$, the local density of states (LDOS) $\rho_\mathbf{i}$, and the local density $n_\mathbf{i}$ (the sum of all sheet populations). To do so,
we generalize the Chebyshev-BdG (CBdG) method \cite{Covaci2010} to determine the real-space time-ordered Nambu-Gor'kov Green's function of a multi-band superconductor. The CBdG method is an efficient numerical method generally applied to inhomogeneous superconductors, where exact diagonalization techniques impose severe restrictions on system's sizes. Here, we take advantage of the CBdG method to investigate homogeneous multi-band superconductors; thereby circumventing the limitations imposed by the size of the matrix Hamiltonian of multi-band materials.
We outline the basic steps below; for more details we refer the reader to reference \cite{Weisse2005b}. 

We write 
the double-time Green's function for band $\nu$ as
\begin{equation}
\label{eq:time_gf}
G_{\mathbf{ij}}^\nu(t,0)=-\frac{i}{\hbar}\begin{bmatrix}
\langle\mathcal{T}c_{\mathbf{i}\nu\uparrow}(t)c^\dag_{\mathbf{j}\nu\uparrow}(0)\rangle & \langle\mathcal{T}c_{\mathbf{i}\nu\uparrow}(t)c_{\mathbf{j}\nu\downarrow}(0)\rangle \\ 
\langle\mathcal{T}c^\dag_{\mathbf{i}\nu\downarrow}(t)c^\dag_{\mathbf{j}\nu\uparrow}(0)\rangle & \langle\mathcal{T}c^\dag_{\mathbf{i}\nu\downarrow}(t)c_{\mathbf{j}\nu\downarrow}(0)\rangle 
\end{bmatrix},
\end{equation}
where $\mathcal{T}$ is the time-ordering operator and $\langle\ldots\rangle$ are thermal-averages. 
One can rewrite equation \eqref{eq:time_gf} with energy arguments as
\begin{equation}
\label{eq:ftransf_gf}
G_{\mathbf{ij}}^\nu(E)=\begin{bmatrix}
\langle c_{\mathbf{i}\nu\uparrow}|G(E)| c^\dag_{\mathbf{j}\nu\uparrow}\rangle & \langle c_{\mathbf{i}\nu\uparrow}|G(E)|c_{\mathbf{j}\nu\downarrow}\rangle \\ 
\langle c^\dag_{\mathbf{i}\nu\downarrow}|G(E)| c^\dag_{\mathbf{j}\nu\uparrow}\rangle & \langle c^\dag_{\mathbf{i}\nu\downarrow}|G(E)|c_{\mathbf{j}\nu\downarrow}\rangle 
\end{bmatrix},
\end{equation}
where
\begin{equation}
G(E)=\lim_{\eta\rightarrow 0}\frac{1}{E-\mathcal{H}+i\eta},
\end{equation}
and $\eta$ is a positive infinitesimal.
For an L$\times$ L square lattice and b bands, the matrix representation of $\mathcal{H}$ has dimension 2bL$^2$. Here we use b=4 and L=30. In our calculations no significant changes were observed for $L>30$.

The diagonal ($\kappa=1$) and off-diagonal ($\kappa=2$) components of equation \eqref{eq:ftransf_gf} -- $G_{\mathbf{ij}}^{\nu,1\kappa}(E)$ -- correspond to the normal and anomalous (superconducting) Green's functions respectively.
In order to expand these components in terms of orthogonal Chebyshev polynomials, we must rescale energy related quantities as $\tilde{E}=(E-b)/a$, where $a=(E_{\mathrm{max}}-E_{\mathrm{min}})/(2-\epsilon)$ and $b=(E_{\mathrm{max}}-E_{\mathrm{min}})/2$, where $\epsilon$ is a small cutoff to avoid stability problems. $E_{\mathrm{max}}$ and $E_{\mathrm{max}}$ are estimates of the bounded extremal values of the eigenvalue spectra of the Hamiltonian. Since we treat a homogeneous system, these estimates can be exactly obtained by diagonalizing a much smaller version of the full $\rm 2{ bL}^2\times 2{ bL}^2$ matrix. The Hamiltonian operator rescales as $\tilde{\mathcal{H}}=(\mathcal{H}-b\mathbb{1})/a$. We indicate all rescaled quantities with a tilde thereafter. 

The components of the Green's function \eqref{eq:ftransf_gf} can be expanded in terms of orthogonal Chebyshev polynomials, which we write as
\begin{equation}
\label{eq:expansion}
\tilde{G}_{\mathbf{ij}}^{\nu,1\kappa}(\tilde{E})=-\frac{i}{\sqrt{1-\tilde{E}^2}}\sum_{n=0}^N \mu_{\mathbf{ij}}^{\nu,1\kappa}(n)\,e^{-in {\rm arccos} (\tilde{E})},
\end{equation}
where the expansion moments are given by
\begin{equation}
\label{eq:dm}
\mu_{\mathbf{ij}}^{\nu,11}(n)=\frac{2}{1+\delta_{n,0}}\langle c_{\mathbf{i}\nu\uparrow}|T_n(\tilde{\mathcal{H}})|c^\dag_{\mathbf{j}\nu\uparrow}\rangle,
\end{equation}
\begin{equation}
\label{eq:odm}
\mu_{\mathbf{ij}}^{\nu,12}(n)=\frac{2}{1+\delta_{n,0}}\langle c^\dag_{\mathbf{i}\nu\downarrow}|T_n(\tilde{\mathcal{H}})|c^\dag_{\mathbf{j}\nu\uparrow}\rangle.
\end{equation}
In this paper only these two components are necessary. We calculate the moments up to expansion order $n=3000$.
The expansion \eqref{eq:expansion} must be convoluted with a proper kernel in order to damp the resulting Gibbs oscillations originating from the Chebyshev polynomials $T_n(\mathcal{\tilde{H}})$. To do so we use the Lorentz kernel, which is designed for Green's functions \cite{Weisse2005b}.
The Chebyshev matrix polynomials $T_n(\tilde{\mathcal{H}})$ obey the recurrence relation
\begin{equation}
\label{eq:rec}
T_{n+1}(\tilde{\mathcal{H}})=2\,\tilde{\mathcal{H}}\,T_{n}(\tilde{\mathcal{H}})-T_{n-1}(\tilde{\mathcal{H}}).
\end{equation}
Using equation \eqref{eq:rec},
the expansion moments \eqref{eq:dm} and \eqref{eq:odm} are obtained by an efficient and stable iterative procedure involving repeated applications of the rescaled Hamiltonian via \eqref{eq:rec} on iterative vectors $|c^\dag_{\mathbf{j}\nu\uparrow}\rangle$.
The diagonal moments \eqref{eq:dm} are used to calculate the local density of states (LDOS)
\begin{equation}
\label{eq:dos}
\tilde{\rho}_{\mathbf{i}}^{\uparrow(\downarrow)}=-\frac{1}{\pi}\sum_{\nu}\mathrm{Im}\,\tilde{G}_{\mathbf{ii}}^{\nu,11(22)}(\tilde{E}).
\end{equation}
For no external magnetic fields $\tilde{\rho}_{\mathbf{i}}^{\uparrow}=\tilde{\rho}_{\mathbf{i}}^{\downarrow}$, such that $\tilde{\rho}_\mathbf{i}=2\tilde{\rho}_{\mathbf{i}}^{\uparrow}$. A final back-scaling yields $\rho_\mathbf{i}$. The local charge density is determined by
\begin{equation}
\label{eq:density}
n_\mathbf{i}=\int_{-1}^1\mathrm{d}\tilde{E}\,\tilde{\rho}_\mathbf{i}(\tilde{E})\tilde{f}(\tilde{E}),
\end{equation}
where the integral is performed in the Chebyshev interval $[-1,1]$ and $\tilde{f}(\tilde{E})$ is the rescaled Fermi distribution. Such integrals can be efficiently calculated using Chebyshev-Gauss techniques \cite{Weisse2005b}.

The off-diagonal moments \eqref{eq:odm} determine the temperature dependence of the real part of the superconducting gaps

\begin{comment}
\begin{equation}
\Delta_\mathbf{i}^\nu(T)=U_\mathbf{i}^\nu\frac{i}{2\pi}\int_{-1}^{1}\mathrm{d}\tilde{E}\,\tilde{G}_{\mathbf{ii}}^{\nu,12}(\tilde{E})\tanh\left(\frac{\tilde{E}}{2\tilde{k}_B T} \right );
\end{equation}
\begin{equation}
\Delta_\mathbf{ij}^\nu(T)=U^\nu_\mathbf{ij}\frac{i}{2\pi}\int_{-1}^{1}\mathrm{d}\tilde{E}\,\tilde{G}_{\mathbf{ij}}^{\nu,12}(\tilde{E})\tanh\left(\frac{\tilde{E}}{2\tilde{k}_B T} \right ).
\end{equation}

In our calculations we are interested only in the real part of the gaps $\mathrm{Re}\,\Delta$. Taking the real part of the gap functions and setting $U\rightarrow -|U|$ we obtain
\end{comment}
\begin{equation}
\label{eq:gaps}
\Delta_\mathbf{i}^\nu(T)=\frac{|U^\nu_\mathbf{i}|}{2\pi}\int_{-1}^{1}\mathrm{d}\tilde{E}\,\mathrm{Im}\,\tilde{G}_{\mathbf{ii}}^{\nu,12}(\tilde{E})\tanh\left(\frac{\tilde{E}}{2\tilde{k}_B T} \right );
\end{equation}

\begin{equation}
\label{eq:gapij}
\Delta_\mathbf{ij}^\nu(T)=\frac{|U^\nu_\mathbf{ij}|}{2\pi}\int_{-1}^{1}\mathrm{d}\tilde{E}\,\mathrm{Im}\,\tilde{G}_{\mathbf{ij}}^{\nu,12}(\tilde{E})\tanh\left(\frac{\tilde{E}}{2\tilde{k}_B T} \right ). 
\end{equation}

\begin{figure}
\includegraphics[width=0.48\textwidth]{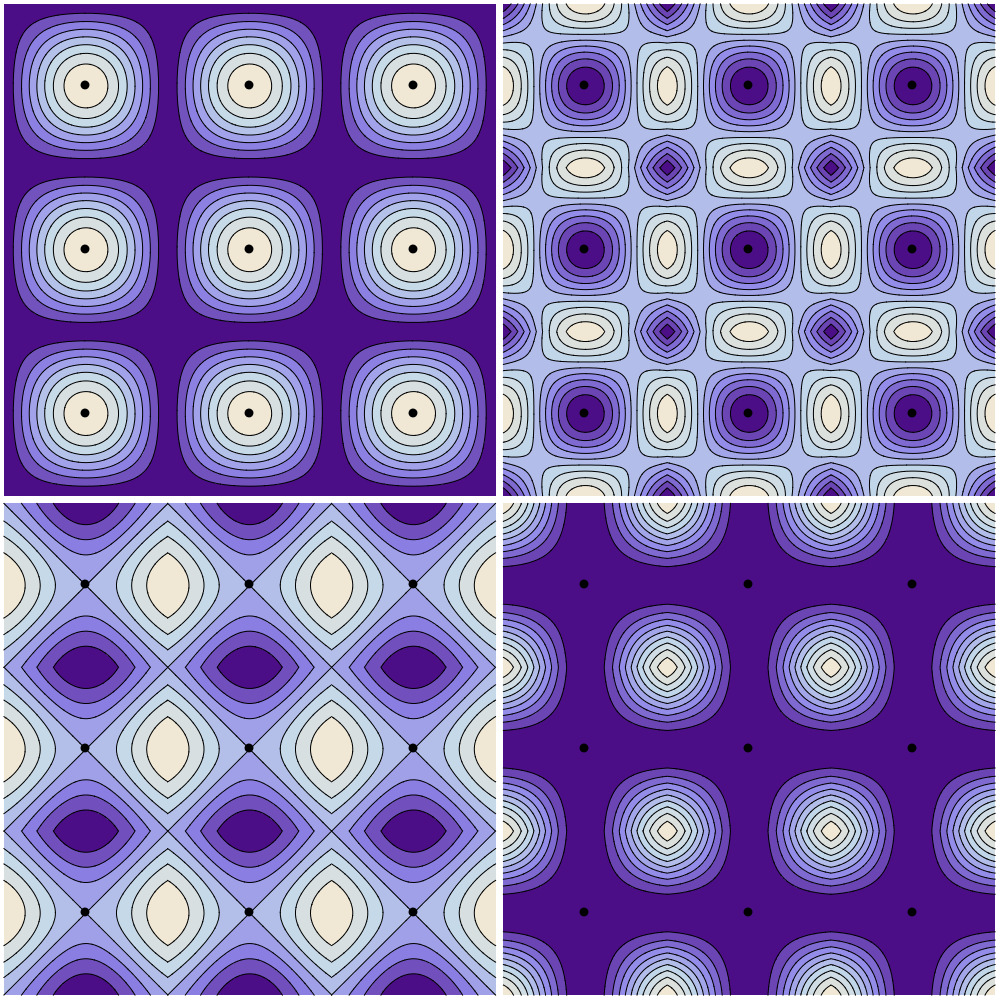}
\caption{\label{fig:realpair} Real space superconducting gap profiles with an underlining square lattice (black dots) correspondent to the gap structures: $s$-wave $\Delta(k_x,k_y)\propto \mathrm{const.}$ (upper-left panel), $s$-wave $\Delta(k_x,k_y)\propto \cos{k_x}+\cos{k_y}$ (upper-right panel), $d$-wave $\Delta(k_x,k_y)\propto \cos{k_x}-\cos{k_y}$ (lower-left panel), and $s$-wave $\Delta(k_x,k_y)\propto \cos{k_x}\cos{k_y}$ (lower-right panel). Lighter colors correspond to higher gap values. The upper-left and lower-right panels show the most likely candidates to describe the FeSCs.}
\end{figure}

For a homogeneous system, we need the list $\{\mu_{\mathbf{ii}}^{\nu,11}(n)\}$ with $\nu=\alpha,\beta,\gamma,\delta$ to calculate the density of states (DOS) from \eqref{eq:dos}, for any $\mathbf{i}$. Similarly, the constant local $s$-wave gap is determined from the list $\{\mu_{\mathbf{ii}}^{\nu,12}(n)\}$, and the nearest neighbor and next-nearest neighbor gaps $\Delta_\mathbf{ij}$ are extracted from the list $\{\mu_{\mathbf{ij}}^{\nu,12}(n)\}$ for any fixed $\mathbf{i}$.

Different combinations of the gap functions \eqref{eq:gapij} emulate a menu of gap symmetries in $\mathbf{k}$-space. In table \ref{tab:latticespace} we show the correspondence of four well-known gap structures in momentum space, with its counterpart in lattice space. To understand the form of the lattice-space combinations, we show the real space profile -- the Fourier transformed momentum gap structures -- with an underlining square lattice in figure  \ref{fig:realpair}. To determine the local unconventional gap value, we take the mean value of the inter-site gap profile \cite{Soininen1994a,Schmid2010}, see table \ref{tab:latticespace}.

\begin{table}
\caption{\label{tab:latticespace}Correspondence between different gap structures in momentum space and lattice space. The vector $\mathbf{x}$ ($\mathbf{y}$) connects two neighboring horizontal (vertical) lattice sites. The vector $\mathbf{d}=\mathbf{x}+\mathbf{y}$ and $\mathbf{g}=\mathbf{y}-\mathbf{x}$.}

\begin{ruledtabular}
\begin{tabular}{ll}
{\bf $\mathbf{k}$-space} & {\bf Lattice space}                                                                                                                                                                               \\
const.                        & $\Delta_{\mathbf{i}}=\Delta_{\mathbf{i}}$                                                                                                                                                         \\
$\cos k_x+\cos k_y$           & $\Delta_{\mathbf{i}}=(\Delta_{\mathbf{i},\mathbf{i}+\mathbf{x}}+\Delta_{\mathbf{i},\mathbf{i}-\mathbf{x}}+\Delta_{\mathbf{i},\mathbf{i}+\mathbf{y}}+\Delta_{\mathbf{i},\mathbf{i}-\mathbf{y}})/4$ \\
$\cos k_x-\cos k_y$           & $\Delta_{\mathbf{i}}=(\Delta_{\mathbf{i},\mathbf{i}+\mathbf{x}}+\Delta_{\mathbf{i},\mathbf{i}-\mathbf{x}}-\Delta_{\mathbf{i},\mathbf{i}+\mathbf{y}}-\Delta_{\mathbf{i},\mathbf{i}-\mathbf{y}})/4$ \\
$\cos k_x\cos k_y$            & $\Delta_{\mathbf{i}}=(\Delta_{\mathbf{i},\mathbf{i}+\mathbf{d}}+\Delta_{\mathbf{i},\mathbf{i}-\mathbf{d}}+\Delta_{\mathbf{i},\mathbf{i}+\mathbf{g}}+\Delta_{\mathbf{i},\mathbf{i}-\mathbf{g}})/4$
\end{tabular}
\end{ruledtabular}
\end{table}

\section{Temperature dependence of the four-gap structure}

%\begin{widetext}
\begin{figure*}
\includegraphics[width=\textwidth]{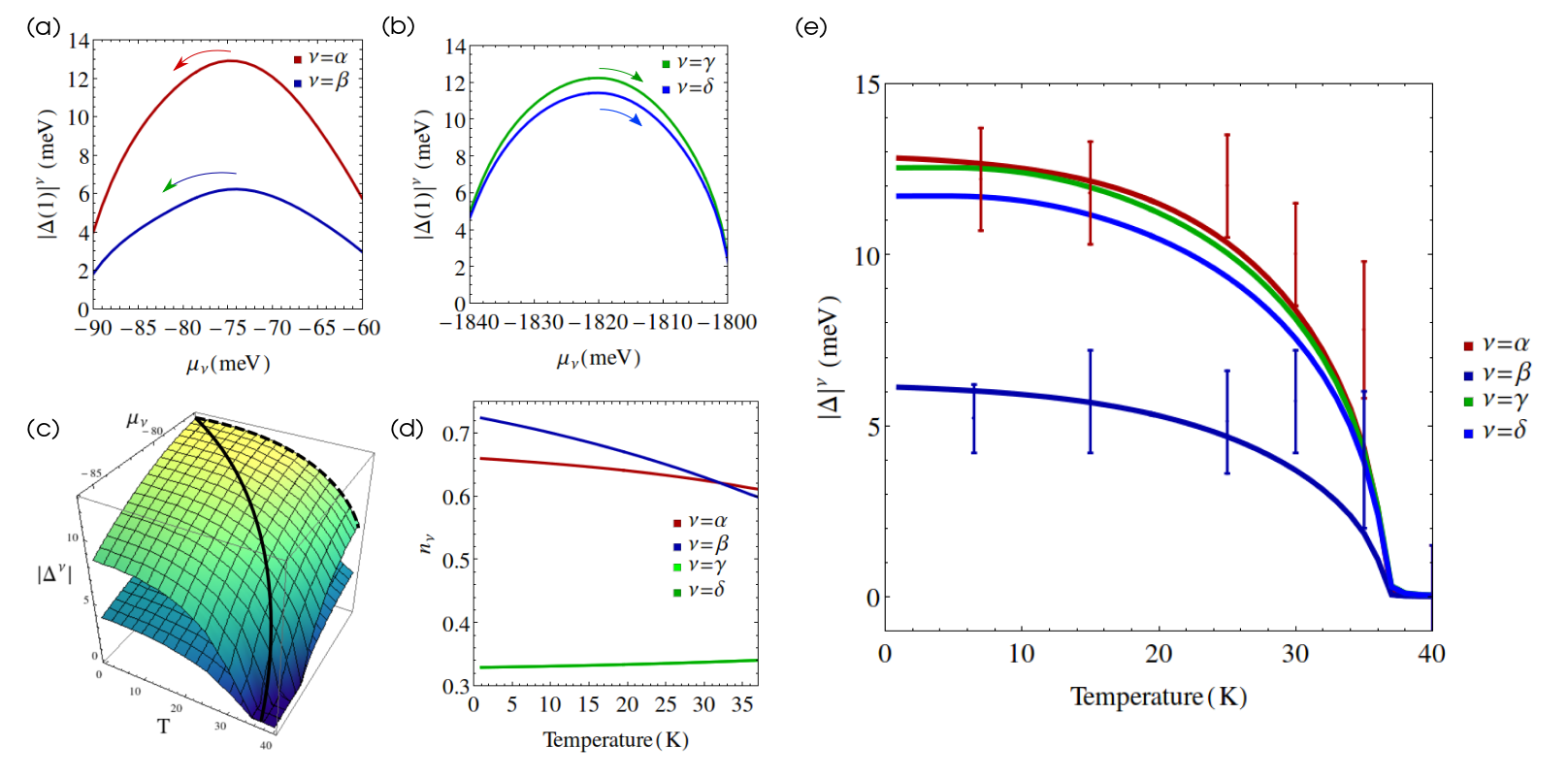}
\caption{\label{fig_bon} (a) Behavior of the superconducting gap in the $\alpha$ and $\beta$ bands with varying chemical potential. 
A monotonic loss (or gain) in intra-band electron population causes higher coupling ratios than the usual BCS result of $2\Delta/k_{\rm B} T\approx 3.52$. The red and the blue arrows indicate the direction of the chemical potential when temperature increases. This implies a monotonic reduction of electron population in the $\alpha$ and $\beta$ bands. See figure c for the 3D version.
(b) The gaps in the $\gamma$ and $\delta$ bands as a function of the chemical potential. To conserve the total electron population, the $\gamma$ and $\delta$ bands gain electrons as indicated by the green and blue arrows. (c) The surface of $\Delta^\alpha(\mu_\alpha,T)$ above with the the surface of $\Delta^\beta(\mu_\beta,T)$ beneath.
The dashed line shows the constant $\mu$ BCS-geodesic $\Delta^\alpha (-76,T)$. The solid line shows a geodesic with varying chemical potential, with initial point at $\Delta^\alpha(-76,1)$ and final point at $\Delta^\alpha(-84.5,37)$, that reproduces the temperature dependence of the energy gaps in FeSCs. Geodesics for the other bands are similar. (d) Redistribution of electron population among the superconducting bands with temperature along the four geodesics. In the present case, $\alpha$ and $\beta$ bands loose electrons, while $\gamma$ and $\delta$ bands gain electrons. The $\gamma$ and $\delta$ curves coincide. (e) Projections of the four geodesics onto the $\Delta^\nu(T)$ plane. We show error bars as extracted from ARPES for the $\alpha$ and $\beta$ band-gaps. The gap in $\gamma$ almost coincides with $\alpha$, and experimental points for the $\delta$-gap were not available.}
\end{figure*}
%\end{widetext}

We examine the case of constant local $s$-wave superconductivity in equation \eqref{eq:gaps}, that is, $U_\mathbf{i}^\nu=\mathrm{const.}$ and all $U_\mathbf{ij}^\nu=0$; see table \ref{tab:parameters}. Weak-coupling BCS multi-band models cannot simultaneously reproduce the experimental values of the superconducting critical temperature and energy gaps, because they yield coupling rations $2\Delta/k_BT_c\approx 3.52$. A BCS-like temperature dependence of the energy gap has constant chemical potential $\Delta^\nu(T)\equiv \Delta^\nu(\mu=\mathrm{const},T)$. This is certainly the case in single-band systems such as the cuprates, where band filling remains constant. However, the FeSCs are intrinsically multi-band systems -- the total electron population $n=\sum_{\nu}n_\nu$ is fixed, but band electron population $n_\nu(T)$ may vary, and hence $\mu_\nu(T)$. 
This opens the possibility that $\Delta^\nu(T)$ may be a geodesic over the surface of $\Delta^\nu(\mu_\nu,T)$, in which the chemical potential $\mu$ is allowed to vary monotonically in a specific band in contrast to the BCS-like constant $\mu$ geodesic over $\Delta^\nu(\mu,T)$, see figure \ref{fig_bon}c. 
To show how the energy gap varies with electron population, we plot $\Delta^\nu(\mu,1)$ in figure \ref{fig_bon}a and figure \ref{fig_bon}b for $\nu=\alpha,\beta$ and $\nu=\beta,\gamma$ respectively.
We call attention to the peaks of the gap intensity in $\alpha$ and $\beta$ bands at $\mu_{\alpha,\beta}=-76$ meV that have correspondent electron populations of $n_\alpha=0.66$ and $n_\beta=0.72$. These band populations can be extracted from the components of equation \eqref{eq:dos}. Simultaneously, the gaps in the $\gamma$ and $\delta$ bands peak at $\mu_{\gamma,\delta}=-1820$ meV, which is consistent with their smaller band population $n_{\gamma,\delta}=0.33$; consistent with pocket sizes as observed by ARPES \cite{Richard2011,Richard2015}.

Our main calculation is explained in figure 2 where the 
calculated  $\Delta^\nu(\mu,T)$ follows a geodesic to left (red and blue arrows in figure 2a) of the peak in the $\alpha$ and $\beta$ hole bands, while $\Delta^\nu(\mu,T)$ in the $\gamma$ and $\delta$ bands follow a geodesic to the right (green and light-blue arrows in figure 2b) on their correspondent peaks. The paths must be different to conserve the total density of the multi-band system.
This allows for high coupling ratios caused by a redistribution of electron population among the bands with varying temperature. While electron population increases in the $\alpha$ and $\beta$ bands with increasing temperature, electron population decreases at the same rate in the $\gamma$ and $\delta$ bands, thus maintaining the total density of the system constant; see figure \ref{fig_bon}d.
%The total density is $\approx 2$, which is consistent with studies that show that two orbitals at nearly half-filling dominate the Fermi surface topology \cite{Hu2012}. 

In figure \ref{fig_bon}c we show the geodesic on the $\Delta^\alpha(\mu,T)$ surface to illustrate the idea, which is analogous for the other surfaces. In figure \ref{fig_bon}e we plot our main results, the projections of the four geodesics on the four $\Delta^\nu(\mu,T)$-surfaces. These projections reproduce the experimental superconducting gap dependence with temperature.
Unfortunately, little experimental data is available for optimally doped Ba$_{1-x}$K$_x$Fe$_2$As$_2$. The experimental error bars we show in figure \ref{fig_bon}e are one of the earliest papers on the temperature dependence of the multi-gap structure in these materials \cite{Ding2008}. 
However, by studying the temperature dependence of the energy gaps for other dopings, it is generally accepted that the formula $\Delta^\nu(T)=\Delta^\nu(0)\tanh(\pi/2\sqrt{T_c/T-1})$ fits the experimental in the FeSCs \cite{Evtushinsky2009a}. Our theoretical CBdG curves coincide exactly with this empirical formula.

We also investigated the unconventional gap structure with $U_\mathbf{i}^\nu=0$ and non-zero second-nearset neigbors $U_\mathbf{ij}^\nu$, which emulates $\cos k_x \cos k_y$ gap symmetry in $\mathbf{k}$-space; see figure \ref{fig:realpair}d. By properly readjusting the $U_\mathbf{ij}^\nu$ and slightly different chemical potentials, one can obtain similar results as obtained by the constant $s$-wave case following the same charge redistribution arguments as discussed above.

\section{Conclusion}

We generalized the CBdG method to treat multi-band superconductors. This allowed us to evaluate $7200\times 7200$ ($ \rm2{ bL}^2\times 2{ bL}^2$) matrices, which would be unfeasible with the exact diagonalization BdG technique. We used the CBdG method's efficiency to address four bands simultaneously, instead of studying inhomogeneous superconductivity. 

We demonstrated that a  multi-band BdG theory can reproduce at a single calculation the high and low coupling ratios $2\Delta/k_{\rm B}T_c$ observed in high-$T_c$ multi-band FeSCs. The
central point of our theory is the SC calculations at the
maxima of $\Delta^\nu(\mu,T)$ with slightly variation of
$\mu(T)$ with the temperature (of the order of 10 - 20 meV).
This represents a small exchange of particle between the overlaping bands $\alpha$ - $\beta$ and $\gamma$ - $\delta$.
The calculated $\Delta^\nu(\mu,T)$ curves were in completely agreement
with the empirical estimates of the experimental results.
Surely these  cannot be used for single-band superconductors such as the cuprates. A multi-band context is indispensable, where band electron populations can redistribute. 

\bibliography{2bands_bib}

\end{document}